\documentclass[sigconf]{acmart}
\usepackage{graphicx} 
\usepackage{soul} 
\usepackage{url}
\usepackage{multirow} 
\usepackage{caption} 
\usepackage{rotating} 
\usepackage{caption} 
\usepackage{tabularx} 
\usepackage{booktabs}
\usepackage{multirow}
\usepackage{glossaries}

\newacronym{ml}{ML}{Machine Learning}
\newacronym{iot}{IoT}{Internet of Things}
\newacronym{nlp}{NLP}{Natural Language Processing}
\newacronym{dl}{DL}{Deep Learning}
\newacronym{dnn}{DNN}{Deep Neural Netwok}
\newacronym{gnn}{GNN}{Graph Neural Netwok}
\newacronym{rnn}{RNN}{Recurrent Neural Network}
\newacronym{cnn}{CNN}{Convolutional Neural Network}

\AtBeginDocument{%
  \providecommand\BibTeX{{%
    \normalfont B\kern-0.5em{\scshape i\kern-0.25em b}\kern-0.8em\TeX}}}

\copyrightyear{2023}
\acmYear{2023}
\setcopyright{acmlicensed}\acmConference[IoT 2023]{The International Conference on the Internet of Things}{November 7--10, 2023}{Nagoya, Japan}
\acmBooktitle{The International Conference on the Internet of Things (IoT 2023), November 7--10, 2023, Nagoya, Japan}
\acmPrice{15.00}
\acmDOI{10.1145/3627050.3627053}
\acmISBN{979-8-4007-0854-1/23/11}

\begin{document}

\title{Enhancing IoT Security via Automatic Network Traffic Analysis: The Transition from Machine Learning to Deep Learning}

\author{Mounia Hamidouche}
\affiliation{%
 \institution{Technology Innovation Institute
}
 \streetaddress{9639 Masdar City}
 \city{Abu Dhabi}
 \country{UAE}}
\email{mounia.hamidouche@tii.ae}

\author{Eugeny Popko}
\affiliation{%
 \institution{Unikie}
 \streetaddress{Hermiankatu 1}
 \city{Tampere}
 \country{Finland}}
\email{eugeny.popko@unikie.com}

\author{Bassem Ouni}
\affiliation{%
 \institution{Technology Innovation Institute}
 \streetaddress{9639 Masdar City}
 \city{Abu Dhabi}
 \country{UAE}}
\email{bassem.ouni@tii.ae}


\begin{abstract}
   This work provides a comparative analysis illustrating how Deep Learning (DL) surpasses Machine Learning (ML) in addressing tasks within Internet of Things (IoT), such as attack classification and device-type identification. Our approach involves training and evaluating a DL model using a range of diverse IoT-related datasets, allowing us to gain valuable insights into how adaptable and practical these models can be when confronted with various IoT configurations. We initially convert the unstructured network traffic data from IoT networks, stored in PCAP files, into images by processing the packet data. This conversion process adapts the data to meet the criteria of DL classification methods. The experiments showcase the ability of DL to surpass the constraints tied to manually engineered features, achieving superior results in attack detection and maintaining comparable outcomes in device-type identification. Additionally, a notable feature extraction time difference becomes evident in the experiments: traditional methods require around 29 milliseconds per data packet, while DL accomplishes the same task in just 2.9 milliseconds. The significant time gap, DL's superior performance, and the recognized limitations of manually engineered features, presents a compelling call to action within the IoT community. This encourages us to shift from exploring new IoT features for each dataset to addressing the challenges of integrating DL into IoT, making it a more efficient solution for real-world IoT scenarios.
\end{abstract}





\maketitle
\section{Introduction}

The proliferation of \gls{iot} applications has increased the importance of evaluating and analyzing \gls{iot} data, primarily to ensure the safety of \gls{iot} systems, which holds the utmost significance \cite{zhang2014iot}. Typically, security concerns related to IoT, including tasks like identifying intrusions, detecting anomalies, and determining device-types, are addressed through three fundamental phases: i) Extracting Features: In this phase, we identify local characteristics or relevant information from \gls{iot} data to represent the data effectively. ii) Selecting Features: We choose the most relevant features from the extracted features during this phase. The goal is to select features that contribute the most to the desired task while reducing redundancy or irrelevant information. iii) Classifying Data: This final phase utilizes the selected features to classify \gls{iot} data into different classes.

\begin{figure*}[t] 
  \centering
  \resizebox{!}{3.431in}{\includegraphics{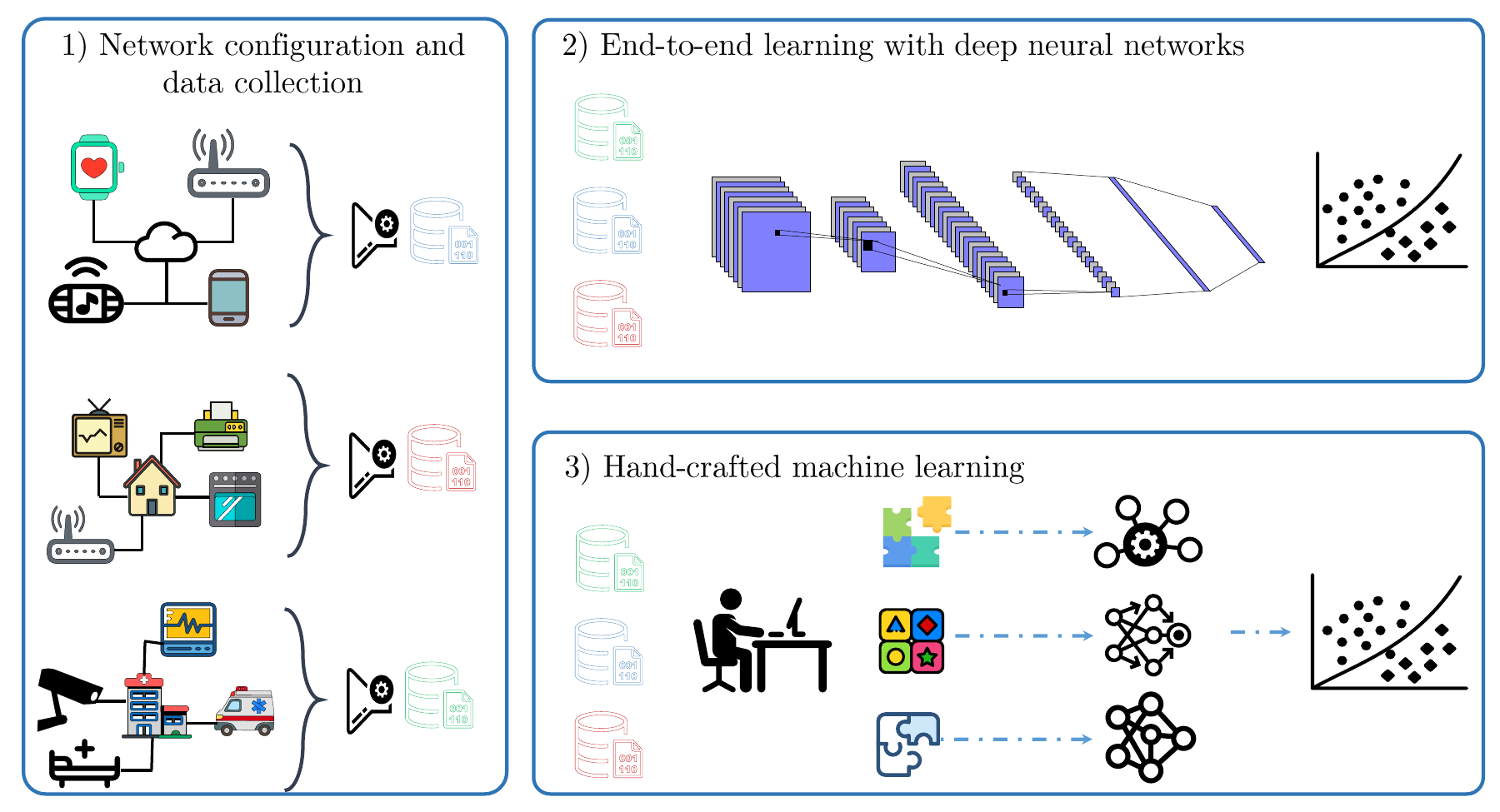}}
  \caption{The figure illustrates a global comparison of hand-crafted method and end-to-end learning using \gls{dnn}for \gls{iot} applications. The first panel shows the diversity of \gls{iot} configurations and the heterogeneous nature of data collected from various networks. In the second panel, we consider a hand-crafted feature learning approach. The engineer analyzes each dataset separately, customizing the features for the specific task. However, these personalized features are not transferable across datasets, making it challenging to represent different datasets effectively. Consequently, this approach involves time-consuming iterations in large-scale data, rendering it impractical. In contrast, the end-to-end learning approach employs a \gls{dnn} model that directly processes the input data, providing a comprehensive solution. This process is easily adaptable to any new dataset, enabling the model to adjust to each dataset's unique characteristics—a capability not inherent in hand-crafted features.}
  \label{fig:your_figure_label}
\end{figure*}

One commonly employed technique for the feature extraction phase in the \gls{iot} domain involves domain specialists' meticulous engineering of features. We craft these features with the specific aim of effectively addressing classification challenges. While this traditional approach is still frequently used due to its interpretability, it possesses inherent limitations: i) Laborious work: Obtaining manually designed features from the data can be time-consuming, involving iterative stages such as experimentation, evaluation, and refinement. This process is repeated for each new dataset and target task because each dataset can showcase distinct characteristics arising from the specificity of the network configuration it originates from (device-type, protocols, and network topologies), as depicted in the first and third panel of Figure \ref{fig:your_figure_label}.  ii) Limited Generalization Evaluation: As per the first limitation, we understand that it becomes difficult to determine how well a \gls{ml} model can extend its classification abilities to different datasets when using a specific set of hand-crafted features. As a result, accurately assessing the model's effectiveness in real-world scenarios, where feature sets may vary, becomes problematic \cite{sarhan2022towards}. iii) Challenges of handling large-scale \gls{iot} Datasets: As discussed in \cite{koc2012network}, manually built features may need help to capture the intricate interactions in large-scale \gls{iot} datasets as handling extensive data volumes can make feature extraction unmanageable and processing computationally impossible.

We aim to go beyond the traditional approach and leverage the potential of end-to-end learning from the \gls{dl} domain to IoT problems, which integrates feature extraction and classification phases into one system, eliminating the need for human intervention, as shown in panel two from Figure \ref{fig:your_figure_label}. This approach draws inspiration from the successes achieved in computer vision \cite{gu2018recent} and \gls{nlp} \cite{mikolov2013efficient}, \cite{mikolov2013distributed} domains, where \gls{dnn} models replaced hand-crafted features because of their outstanding performance.
The adaptability of end-to-end learning allows the model to seamlessly work with diverse \gls{iot} datasets and tasks, significantly reducing time consumption and laborious effort, addressing the previously mentioned limitation i. Furthermore, by learning directly from raw data, the model gains a deeper understanding of the underlying patterns and relationships present in the data. This enhanced understanding allows the model to grasp unique feature sets and adapt more effectively to different scenarios and datasets, addressing the previously mentioned limitation ii. Moreover, end-to-end learning is particularly advantageous for large-scale \gls{iot} datasets, enhancing scalability and computational efficiency, addressing limitation iii.

Our study uses an end-to-end approach for two vital \gls{iot} tasks: Attack detection (binary and multi-class) and device-type recognition with three classes. Both tasks are crucial for improving \gls{iot} security. They are interconnected because device-type recognition can aid attack detection, allowing the efficient assessment of security threats related to different device vulnerabilities \cite{miettinen2017iot}.

Typically, organizations store network traffic data in unstructured files like PCAP files. These files contain a collection of packets, each representing a discrete unit of data transmitted over the network. Each packet in the PCAP file typically includes header information such as the source and destination IP addresses, port numbers, packet timestamp, and the actual content of the packet. This data needs to be more structured for traditional end-to-end learning methods. To tackle this challenge and leverage the power of \gls{dl} algorithms for our two tasks, we implement a contemporary approach outlined by \cite{wang2017malware}, originally designed for a different purpose, namely, malware traffic classification, but we adapt it for our purposes. Using this technique, we successfully transformed the unstructured data within PCAP files into an image-based representation and used \gls{dl} to detect various attacks and device-types. 

We carried out experiments on a total of four datasets. Among these, one dataset is employed for attack identification \cite{CICC}. This dataset comprises three types of classifiers: a binary classifier to assert the presence or absence of an attack, an 8-class classifier to categorize the attacks into various attack types, and a 34-class classifier to identify the specific attack precisely, see Table \ref{Table:attacks}. We utilized three datasets for device-type identification task, namely, UNSW-Sydney 2018 (TMC) \cite{UNSW-Sydney}, CIC IoT 2022 (CIC22) \cite{dadkhah2022towards}, and IMC 2019 (IMC) \cite{ren2019information}. In these instances, the primary goal is to classify devices into three distinct categories: audio devices, cameras, and home automation, as illustrated in Table \ref{Table:devices}.

Our experiments show that end-to-end learning method exhibits high quality as it achieves comparable, and in some cases, even superior results compared to the outcomes obtained through manual feature engineering. Moreover, the experimental results indicate that using a single \gls{dl} model enables us to successfully tackle multiple tasks with different datasets. In contrast, when researchers attempted to address the same functions and datasets using other \gls{ml} methods and hand-crafted features, they had to identify and select the best features for each specific combination of dataset and task, as each dataset had its unique set of features. Furthermore, we noted a substantial disparity between these two approaches concerning the time required for feature extraction. Hand-crafted features, which have demonstrated effectiveness in device identification, exhibited a relatively lengthy processing time of approximately 26 milliseconds for each data packet. In contrast, \gls{dl} achieved the same task with remarkable efficiency, completing it in just 0.3 milliseconds. This substantial contrast in processing time underscores the practicality of \gls{dl} in real-world applications. 

The structure of this paper is as follows: Section 2 provides an overview of related work and discusses the motivation behind our research. Section 3 presents the end-to-end learning-based methodology employed for device-type identification and attack detection tasks in the \gls{iot} domain. Section 4 presents the results of our experiments. We conduct a comparative analysis between models utilizing hand-crafted features and those employing learned feature extraction. We discuss the limitations of our current method and outline potential directions for future work in Section 5. Finally, in Section 6, we provide concluding remarks summarizing the contributions of our study. We make the setup code, dataset, data pre-processing, and detection model available to the public through GitHub~\footnote{\href { https://owl-light-racer.ngrok-free.app/mounia.hamidouche/iot-securedl}{"
You can access the scripts and datasets by visiting the following URL: https://owl-light-racer.ngrok-free.app/mounia.hamidouche/iot-securedl"} \label{source_code}}.

\section{Related Work}

This section overviews relevant research on intrusion detection and device identification in \gls{iot} systems. Previous studies predominantly rely on supervised \gls{ml} algorithms with hand-crafted features for data representation, utilizing techniques such as decision trees, rule-based systems, neural networks, support vector machines, naive Bayes, and k-nearest neighbor for classification. Many works in device identification and attack detection evaluate their \gls{ml} models using benchmark datasets due to the challenges in obtaining realistic labeled network traffic for testing. These evaluation methods aim to identify practical features and ML models for the classification.

Notable research in device-type identification includes \cite{marchal2019audi}, \cite{meidan2017profiliot}, \cite{meidan2017detection}, \cite{shahid2018iot}, \cite{aksoy2019automated}, \cite{chowdhury2020network}, \cite{sivanathan2018classifying}, and \cite{mainuddin2022iot}. Similarly, for the task related to attack detection, we find various works \cite{kasongo2020performance},  \cite{koroniotis2019towards}, \cite{moustafa10ton}, \cite{CICC}, and many others. These studies use traffic properties like network flow, protocols, and TCP sessions as features to differentiate between \gls{iot} and non-\gls{iot} devices or classify various types of attacks. While the mentioned studies contribute significantly to device identification and intrusion detection, it is essential to note that these studies typically involve customizing features tailored to each dataset. For instance, datasets such as UNSW-NB15 \cite{kasongo2020performance}, BoT-\gls{iot} \cite{koroniotis2019towards}, ToN-\gls{iot} \cite{moustafa10ton}, and CIC IoT 2023 \cite{CICC} are furnished with a tailored set of features, carefully crafted to attain the best possible detection performance. As the landscape of IoT datasets continues to expand in terms of scale, complexity, and diverse IoT configurations, researchers often find themselves adapting features to accommodate the unique characteristics of their datasets. While this customization is valuable for addressing dataset-specific nuances, it has resulted in a fragmentation of feature sets across different studies. Consequently, this fragmentation poses a challenge when comparing and generalizing to new data due to the heterogeneous nature of \gls{iot} data. The authors in \cite{sarhan2022towards}, \cite{booij2021ton_iot} provide further details, revealing significant differences among the feature sets.

Adopting an end-to-end learning approach offers a promising solution. Unlike hand-crafted features tailored to specific datasets and tasks, end-to-end learning is a data-driven approach that allows the model to automatically learn and adapt to various datasets and tasks, as depicted in Figure \ref{fig:your_figure_label}. As a result, researchers have shifted their focus to \gls{dnn} methods for \gls{iot} problems. For example, in \cite{pujol2022unveiling}, the authors propose a novel \gls{gnn} model that learns from graph-structured information, including flow records and their relationships, instead of relying solely on flow records. Similarly, other studies, such as \cite{yang2022transfer}, \cite{de2021machine}, \cite{basnet2019towards}, \cite{wang2023iot}, and \cite{yang2022enhanced}, employ \gls{dnn} for \gls{iot} tasks. In \cite{yin2017deep}, the authors investigated the use of \gls{dl} for intrusion detection and proposed the \gls{rnn} intrusion detection in binary and multiclass classification tasks. However, these studies employ manually selected features like protocol-type, service, and flag as input variables for the \gls{dl} models. While \gls{dl} has the potential to extract more meaningful representations and create better models, relying on hand-crafted features as inputs may limit its effectiveness, as it might initially miss some important features.

Recently, there has been a notable surge in interest in implementing automatic feature extraction techniques in IoT. With the ever-expanding volume of data, researchers have been actively exploring the potential of utilizing computer vision and \gls{nlp} methods within the IoT domain. However, the IoT domain presents a challenge due to the unstructured nature of the data. To address this challenge, researchers propose innovative techniques. For instance, in \cite{wang2017malware}, \cite{wang2017end},\cite{farrukh2022payload}, the authors introduce novel methods to convert packets from PCAP files into images to use data representation techniques from the computer vision field for malware traffic classification. Similarly, \gls{nlp} techniques are employed to extract features from textual descriptions like log files, as demonstrated in \cite{lotfollahi2020deep}, \cite{ferriyan2022encrypted}, \cite{goodman2020packet2vec},\cite{lu2021iclstm}. In 2020, the authors in \cite{jeon2020malware} introduced a DL-based malware detection system. They employed the convolutional encoder to translate the opcode sequences extracted from Windows executable files, followed by utilizing \gls{rnn} for the malware detection process.  

The research in \cite{wang2017end} and \cite{farrukh2022payload} influences our work, introducing a method for converting network traffic packets into images. By expending this innovative approach, we utilize a single \gls{dl} model that autonomously learns features, resulting in significant time savings. DL models can extract relevant features directly from the data, thereby eliminating the need for laborious feature engineering. This advantage becomes particularly evident when dealing with diverse datasets and tasks, as the DL model adapts and generalizes across various scenarios, streamlining the overall research process. By leveraging the power of DL techniques, we aim to reduce the time and effort previously spent on feature engineering, thereby enabling researchers to shift their focus towards addressing the multifaceted challenges that arise when implementing DL in the context of the IoT. It is essential to recognize that integrating DL into IoT introduces a unique set of hurdles. Interpretability is a challenge, particularly crucial in IoT scenarios where understanding the rationale behind DL model decisions is vital. Deploying DL models on resource-constrained IoT devices entails challenges in computation, energy efficiency, and real-time processing. Scalability, data privacy, adaptability to dynamic environments, and resilience to environmental factors demand meticulous attention and innovative solutions to fully harness DL's potential in IoT applications.

\section{Methodology}
In this section, we describe the methodology employed for designing an end-to-end system responsible for attack detection and device-type identification.  It comprises three key stages: 1) PCAP to image conversion (PCAP parsing), 2) Image preprocessing, and 3) Image classification using \gls{dnn}.

\subsection{PCAP Parsing}
The initial phase involves using the technique from \cite{wang2017end} and \cite{wang2017malware} to transform PCAP files into images. This transformation enables the utilization of \gls{dl} techniques to deal with unstructured data from our \gls{iot} datasets, thereby facilitating the classification process. This stage is divided into three distinct steps. 

\subsubsection{Traffic split}
The raw traffic needs to be split into discrete units. The most common two choices of traffic representation are session and flow \cite{dainotti2012issues}. A session is a traffic unit divided based on 5-tuple, i.e. source IP, source port, destination IP, destination port and transport-level protocol. A flow is very
similar to a session, and the difference is that it contains traffic of only one direction, i.e. the source and destination IP/port are not interchangeable. The traffic byte data in each packet can be divided into multiple protocol layers. There are two types of layer choice in our work. The first choice is layer 7 in ISO/OSI model or layer 4 in TCP/IP model (L7). Intuitively the characteristics of traffic should be reflected in this layer. For example, STMP protocol represents email traffic and HTTP protocol represents browser traffic. The input data for this step is in the PCAP format. There are are four possible representation: 1) "Flow + All", 2) "Session + All", 3) "Flow + L7" and 4) "Session + L7". Throughout our experiments, we employed the "Session + All" format.

\subsubsection{Traffic clearing and length trimming}
This step involves clearing the obtained traffic from Step 1 and fixing the packet length. Some packets generate identical files with the same content, and duplicated data can introduce bias when training CNN. We remove these empty and duplicated files. We also transform all files to have a uniform length. It should be noted that different flows or sessions may have different size, but the input data size of CNN must be uniform, so only first n bytes (n = 784 in
this work) of each flow or session are used. If the file size exceeds 784 bytes, trim it down to 784 bytes. If the file size is less than 784 bytes, add 0x00 to complement it to 784 bytes.
\subsubsection{Image generation}
The output files obtained from the previous step, all possessing the same size, are transformed into grayscale images (23 $\times$ 23). In this conversion process, every byte within the PCAP files serves as a pixel, and these pixels have values spanning from 0 to 255. To accomplish this, we employ a specialized function known as getMatrixfrom\_pcap. This function helps to transform the packet data into structured data (image representation). It is important to note that, during device-type identification, we use a method that transforms each packet within the PCAP file into an image representation. However, given the large dataset involved in the attack detection task, we opt for a more efficient approach. Specifically, we create sequences of images, meaning that we take a sequence representative of each class and convert it into a single image, which helps streamline the processing time.

\subsection{Image Preprocessing}
During this stage, images are normalized and converted to tensors for neural network input by scaling pixel values to [0, 1].  We perform this step before training to ensure the data's proper preparation for practical neural network training.

\subsection{\gls{dnn} Classifier}
Here we implement the \gls{dnn} classifier. The neural network receives the preprocessed image tensors as input and trains to classify the images into their respective categories for either attack detection or device-type identification tasks. Throughout our experiments, the images cover three distinct scenarios in the attack detection task: binary, 8-class, and 34-class classifications. On the other hand, for device-type identification, the classifier deals with three classes.


\section{Experiments}
In this section, we evaluate the effectiveness of our approach on two distinct tasks. The first task involves attack classification, using a dataset containing benign and attack packets. The second task pertains to device-type classification, for which we utilize three separate datasets containing only benign data. In Table \ref{Table:dataset_info}, we provide a general summary of each dataset used in our analysis, presenting information on the capture year, device count, and the data type in each dataset. In the following subsections, we provide detailed descriptions of the datasets employed, the \gls{dnn} architecture utilized, and an extensive comparison of our approach's performance against state-of-the-art results.

\begin{table}[h]
\small
\centering
\caption{Dataset Overview: Device Count (DC), Capture Year (CY), Data Type (DT).}
\begin{tabular}{lcccc}
\toprule
\textbf{Dataset} & \textbf{CY} & \textbf{DC} & \textbf{DT}\\
\midrule
\textbf{TMC} & 2018 & 31 & Benign \\
\textbf{CIC22} & 2022 & 60 & Benign \\
\textbf{IMC} & 2019 & 81 & Benign\\
\bottomrule
\textbf{CIC23} & 2023 & 105 & Benign+Attack\\
\end{tabular}
\label{Table:dataset_info}
\end{table}

\subsection{Dataset Presentation}

For our analysis, we employ three datasets for device-type identification and one for attack detection.

\subsubsection{UNSW-Sydney 2018 (TMC) \cite{sivanathan2018classifying}, \cite{UNSW-Sydney}} This dataset comprises data from 28 \gls{iot} devices placed in a living lab to replicate an intelligent environment. The authors collected and combined the data over a six-month period.

\subsubsection{IMC 2019 (IMC) \cite{ren2019information}} This dataset includes 81 \gls{iot} devices that the authors collected from diverse sites. Among them, the authors purchased 46 devices in the US and installed them in a US testbed, while they bought 35 devices in the UK and installed them in a UK testbed. Notably, 26 devices are common to both testbeds, indicating the presence of identical model names in both locations. In our study, we focused solely on devices from the UK testbed.

\subsubsection{CIC IoT 2022 (CIC22) \cite{dadkhah2022towards}} This dataset is from the Canadian Institute for Cybersecurity (CIC). It contains information collected from 60 \gls{iot} devices deployed in the authors' lab. 

Each of these datasets categorizes the devices into three distinct classes: audio, cameras, and home automation, based on the types of devices present. A comprehensive table with detailed information about the devices included in each dataset is provided in Table \ref{Table:devices}.

\subsubsection{CIC \gls{iot} 2023 (CIC23)} The Canadian Institute for Cybersecurity released this recent dataset \cite{CICC}, \cite{CIC23}. Its specific design encourages the development of security analytics applications for real-world \gls{iot} operations. It presents a novel and extensive collection of \gls{iot} attack data. It includes 33 attacks executed within an \gls{iot} topology comprising 105 devices. The author categorized the attacks into seven classes: DDoS, DoS, Recon, Web-based, Brute Force, Spoofing, and Mirai. In Table \ref{Table:attacks} we provide detailed information about all the attacks we considered and the number of packets used for each attack during our analysis.


\begin{table}[h]
\small
\centering
\caption{ CIC23 IoT Dataset.}
\begin{tabular}{l|c|c}
\toprule
\textbf{Attack Category} & \textbf{Attack} & \textbf{Packet Number} \\
\midrule
\multirow{12}{*}{DDoS} & ACK Fragmentation & 285,104  \\
& UDP Flood & 5,412,287 \\
& SlowLoris & 23,426 \\
& ICMP Flood & 7,200,504  \\
& RSTFIN Flood & 4,045,285  \\
& PSHACK Flood & 4,094,755  \\
& HTTP Flood & 28,790  \\
& UDP Fragmentation & 286,925  \\
& ICMP Fragmentation & 452,489   \\
& TCP Flood & 4,497,667   \\
& SYN Flood & 4,059,190  \\
& SynonymousIP Flood & 3,598,138  \\
\midrule
\multirow{4}{*}{DoS} & TCP Flood & 2,671,445  \\
& HTTP Flood & 71,864  \\
& SYN Flood & 2,028,834 \\
& UDP Flood & 3,318,595\\
\midrule
\multirow{4}{*}{Recon} & Ping Sweep & 2262 \\
& OS Scan & 98,259 \\
& Vulnerability Scan & 37,382  \\
& Port Scan & 82,284  \\
& Host Discovery & 134,378 \\
\midrule
\multirow{5}{*}{Web-Based} & Sql Injection & 5245 \\
& Command Injection & 5409 \\
& Backdoor Malware & 3218  \\
& Uploading Attack & 1252  \\
& XSS & 3846  \\
& Browser Hijacking & 5859 \\
\midrule
\multirow{1}{*}{Brute Force} & Dictionary Brute Force & 13,064  \\
\midrule
\multirow{2}{*}{Spoofing}& Arp Spoofing & 307,593  \\
& DNS Spoofing & 178,911 \\
\midrule
\multirow{3}{*}{Mirai} & GREIP Flood & 751,682  \\
& Greeth Flood & 991,866  \\
& UDPPlain & 890,576  \\
\bottomrule
\end{tabular}
\label{Table:attacks}
\end{table}

\begin{table*}[t]
\small
\centering
\caption{Devices Used for Device-type Identification.}
\begin{tabular}{lcccc}
\toprule
\textbf{Device-type} & \textbf{CIC22} & \textbf{IMC} & \textbf{TMC} \\
\midrule
\multirow{4}{*}{\centering Audio} & Amazon Echo Dot, Amazon Echo Spot  & Echodot, EchoPlus&  Amazon Echo\\
& Amazon Echo Studio, Amcrest &  Google home, EchSpot   & Triby Speaker \\
& Google Nest Mini, Sonos One Speaker & & \\
\midrule
\multirow{4}{*}{\centering Camera} & Nest, SimCam, Dlink, ArloQ  & DXiomiCam, bosiwo-wifi   & Netatmo Welcome,  Dropcam\\
&  Netatmo, HeimVision, Borun, Luohe & blink security-hub & TP-Link Day Night Cloud camera \\
& Arlo Base Station, Home Eye& delinkCam, Blink & Samsung SmartCam, Nest Dropcam  \\ 
\midrule
\multirow{4}{*}{\shortstack{Home\\Automation}}  
& Amazon Plug, SmartBoard, Roomba Vacuum&  Nest-tstat & Smart Things, PIX-STAR Photo-frame  \\
& Globe Lamp, Ring Basestation, Philips Hue Bridge&  tplink-bulb & Belkin Wemo switch \\
& Coffee MakerStation, HeimVision Lamp, Yutron plug&  tplink-plug&  TP-Link Smart plug, NEST Protect smoke alarm \\
& Gosund Plug, Eufy HomeBase, Tekin Plug& t-wemo-plug & Belkin wemo motion sensoR  \\
\bottomrule
\end{tabular}
\label{Table:devices}
\end{table*}

\subsection{\gls{dnn} Architecture}
Researchers developed various neural network architectures to address different problems. Among these architectures, \gls{cnn} models gained significant popularity due to their strong generalizability and impressive performance in image classification tasks. In this study, we opt for a neural network structure similar to LeNet-5 \cite{lecun1998grad}. LeNet-5 is a classic CNN architecture effective in image classification tasks. The architecture of our model comprises the following layers, as illustrated in Table \ref{Table:dataset_mutation_size}:

1. \textbf{Convolution Layer}. The convolution layer is the key component of CNN models. The convolution operation extracts different features of the input~\cite{lecun1998grad}. 
This layer consists of a set of learnable parameters known as filters or kernels. The first layer is a convolution layer with 32 feature maps and a $5\times 5$ kernel for each feature map, and the next layer has 64 feature maps and a $5\times 5$ kernel for each feature map.

2. \textbf{ReLU Activation Layer}. The ReLU layer transforms the previous layer's output using the relu(v) = max(v, 0) function. We employ ReLU as the activation function, delivering comparable or superior results to sigmoid-like functions while converging faster.

3. \textbf{Max Pooling Layer}. Pooling means the downsampling of an image. This study uses Max Pooling with $2\times 2$ filters and stride 2.

4. \textbf{Fully-connected Layer}. This layer is a layer that has interconnected neurons. The input of the first fully connected layer is 64 feature graphs, and their size is $5\times 5$, and the output is
512. The input of the second fully connected layer is 512, and the number of its output neurons reflects the number of data classes.

Our model requires modest computational resources, making it suitable for real-world applications. The neural network architecture employed in this task consists of N output neurons at the top layer. N varies considering the appropriate number of classes for each particular scenario.
\begin{table}[h]
\small
\centering
\caption{CNN Model Architecture.}
\begin{tabular}{lcccc}
\toprule
\textbf{Layer} & \textbf{Layer Type} & \textbf{Size} & \textbf{Output}\\
\midrule
\textbf{1} & Conv + ReLU & 32, $5\times5$, stride: 1 & $32\times 28\times 28$ \\
\textbf{1} & Max Pooling & $2\times 2$, stride: 2 & $32\times 14\times 14$ \\
\textbf{2} & Conv + ReLU & 32, $5\times 5$, stride: 1 & $64\times 10\times 10$ \\
\textbf{2} & Max Pooling & $2\times 2$, stride: 2 & $64\times 5\times 5$ \\
\textbf{3} & Fully Connected + ReLU & 1600 hidden units & 512 \\
\textbf{4} & Fully Connected & 512 hidden units & N classes \\
\end{tabular}
\label{Table:dataset_mutation_size}
\end{table}

\subsection{DNN Training}
We shuffle each dataset to ensure randomness during training. Then, we split it into three sets: 80\% for training, 15\% for validation, and 5\% for testing. This division allows us to train the model on most of the data, validate its performance on a separate set, and test its generalization on unseen data. We used the Adam optimization algorithm, known for its adaptive learning rate and efficient handling of sparse gradients. We set the specific parameters for Adam: a learning rate of 0.001, a first-moment exponential decay rate (beta1) of 0.9, a second-moment exponential decay rate (beta2) of 0.999, and an epsilon value of 1e-8. Epsilon is a small constant added to the denominator to prevent division by zero during calculations. We use a weighted cross-entropy loss function to address the class imbalance issues in the image dataset. This loss function gives more importance to underrepresented classes, allowing the model to focus on improving its performance on the minority classes. The training process consists of 40 epochs, meaning the model underwent 40 complete passes over the training dataset. We use a batch size of 256, allowing the model to process 256 samples simultaneously in each iteration.  We follow this process using Python and PyTorch to train the model on each dataset. We conducted our experiments on a server with the following hardware configuration: AMD EPYC 74F3 24-Core 3.2 GHz CPU, 128 GB of RAM, and an NVIDIA A100 GPU with 80 GB of memory.

\subsection{Defining a Standard Set of Hand-crafted Features for Comparative Analysis}
We use a set of features that have demonstrated promising results in prior research efforts to conduct a comparison with our end-to-end approach in terms of both accuracy and the time it takes to extract these features \cite{erfani2021feature}, \cite{CICC}. This comparative evaluation will enable us to gauge the performance and efficiency of our system when compared to established feature-based methods. After cleaning the data, we provide a list of features we kept in Table \ref{Table:features}. For the description of each feature, please refer to \cite{CICC}.

\begin{table}[t]
\small
\centering
\caption{Selected Set of Hand-crafted Features \cite{CICC}.}
\begin{tabular}{llcc}
\toprule
& \multicolumn{1}{c}{\textbf{Hand-crafted Features}} \\
\midrule
& Variance, AVG, Flow\_active\_time, Drate, TCP, Flow\_idle\_time, Min \\
& Sequence\_number, Header\_Length, Wifi\_src, Rate, LLC,  Max\\
& MAC, Ack\_count, Tot\_size, SSH, Source\_port, TNP\_per\_proto\_tcp \\ 
& Syn\_flag\_number, Urg\_count, Number, Urg\_flag\_number, ICMP  \\
& HTTP, Covariance, RARP, SMTP, Std, Wifi\_Type, MQTT, Sbytes\\
& DS\_status, HTTPS, Srate, Ack\_flag\_number, IRC, Dst\_ip\_bytes \\
& Protocol\_version, UDP, Correlation, Rst\_flag\_number, DHCP, IAT \\
& Dpkts, Rst\_count, IGMP,  Src\_ip\_bytes, Ts, DNS, CoAP, Telnet  \\
& Src\_pkts, ARP, Protocol\_type, IPv, Magnitue, Std\_flow\_duration \\
& Destination\_port, Weight, Average\_flow\_duration, Fin\_count \\
& Spkts, Dst\_pkts, Fin\_flag\_number, Fragments, Sum\_flow\_duration  \\
& Min\_flow\_duration, Cwr\_flag\_number, Tot-sum, Syn\_count\\
& Ece\_flag\_number, Psh\_flag\_number, Max\_flow\_duration \\
\bottomrule
\end{tabular}
\label{Table:features}
\end{table}

\subsection{Evaluation Metrics}
In our study, we use four metrics for the performance evaluation. The metrics definitions are listed below: accuracy (ACC), precision (PRE), recall (REC), and F1-score (F1). We use accuracy to evaluate the overall performance of a classifier. Precision, recall, and F1-scores to assess the performance of every traffic class. 
\begin{equation}
    accuracy = \dfrac{TP+TN}{TP+FP+FN+TN}, \ \ \ \ precision = \dfrac{TP}{TP+FP}
\end{equation}

\begin{equation}
    recall = \dfrac{TP}{TP+FN}, \ \ \ \ F1 \ score = \dfrac{2. precision . recall}{precision + recall}.
\end{equation}

Where true positive (TP) is the number of instances correctly classified as X, true negative (TN) is the number of instances correctly classified as Not-x, false positive (FP) is the number of instances incorrectly classified as X, and false negative (FN) is the number of instances incorrectly classified as Not-X. 

\subsection{Performance Comparison with State-of-the-Art}
In this section we present the results obtained for the two tasks using our approach and compare them with works that utilize hand-crafted features for the same datasets. Specifically, for the attack detection problem, we compare our results to those obtained in \cite{CICC}. While the mentioned paper explored various \gls{ml} algorithms, we specifically focus on comparing our results to their outcomes using random forest (RF), as it demonstrates the best performance. To ensure a fair and meaningful comparison with the work by \cite{CICC}, we replicated their analysis by employing the same hand-crafted features on a subset of the dataset similar to the one we used to train our model. We took this step to ensure fairness and avoid any biases that might arise from using different sizes of datasets. Our comparison is limited to this recent paper since, to our knowledge, they are the sole researchers who have analyzed this dataset.

\subsubsection{Attack Detection}
In the binary classification scenario, our proposed model achieved high performance with: 99.71\% accuracy, 98.89\% recall, 98.05\% precision, and an F1-score of 98.47\%. The results obtained using hand-crafted features yielded high accuracy at 99.68\%, but showed slightly lower recall 96.51\%, precision 96.54\%, and F1-score 96.52\%. For the 8-class classification scenario, similarly, our model and the state-of-the-art model exhibited high accuracy, with our approach achieving 98.76\% and the hand-crafted method obtaining 99.43\%. However, our model excelled in other crucial measures, with recall at 96.81\%, precision at 97.50\%, and a noteworthy F1-score of 97.15\%. In contrast, the hand-crafted features approach resulted in 91.00\% recall, 70.54\% precision, and an F1-score of 71.92\%. The comparison between the two methods indicated that the accuracy scores were closely similar for both binary and 8-class scenarios, suggesting that both approaches achieved a comparable level of accurate classifications. Yet, a closer look at performance metrics (recall, precision, F1-score) revealed a clear advantage for the end-to-end method, which displayed markedly superior recall rates, precision levels, and overall F1-scores compared to the hand-crafted approach. In the most challenging 34-class classification scenario, our model outperformed the hand-crafted feature approach. We achieved 98.81\% accuracy, 95.13\% recall, 92.01\% precision, and an F1-score of 93.28\%. In comparison, the hand-crafted features approach resulted in 99.16\% accuracy, 83.15\% recall, 70.44\% precision, and an F1-score of 71.00\%. We present our system's performance on attack detection in Figure \ref{Table:classification_attack}.

\begin{table}[h]
\small
\centering
\caption{Comparative Analysis of Attack Classification on the CIC23 dataset: Hand-Crafted Features vs. End-to-End Learning.}
\begin{tabular}{c|c|c|c|c|c}
\toprule
\textbf{Features} & \textbf{State} & \textbf{ACC (\%)} & \textbf{REC (\%)} & \textbf{PRE (\%)} & \textbf{F1 (\%)} \\
\midrule
\multirow{3}{*}{\rotatebox[origin=c]{0}{\begin{tabular}[c]{@{}c@{}} \textbf{Automatic} \\ \textbf{Features}\end{tabular}}} & Binary & \textbf{99.71} & \textbf{98.89} & \textbf{98.05} & \textbf{98.47} \\
& 8-class & 98.76 & \textbf{96.81} & \textbf{97.50} & \textbf{97.15} \\
& 34-class & 98.81 & \textbf{95.13} & \textbf{92.01} & \textbf{93.28} \\
\midrule
\multirow{3}{*}{\rotatebox[origin=c]{0}{\begin{tabular}[c]{@{}c@{}} \textbf{Hand-crafted}\\ \textbf{Features} \end{tabular}}} & Binary & 99.68 & 96.51 & 96.54 & 96.52 \\
& 8-class & 99.43 & 91.00 & 70.54 & 71.92 \\
& 34-class & 99.16 & 83.15 & 70.44 & 71.40 \\
\bottomrule
\end{tabular}
\label{Table:classification_attack}
\end{table}

\subsubsection{Device-type Identification}

In the results of our experiments, we observed consistently high accuracy scores, as well as excellent recall, precision, and F2 score values, all of which approached 100\%. These results reflect the remarkable performance of our model in various aspects of classification. Importantly, they are highly comparable those achieved through manual hand-crafted feature engineering. However, a notable advantage of our automated approach is its adaptability. Unlike hand-crafted features, which demand significant effort and expertise for adaptation to complex datasets, our model inherently adjusts.It autonomously extracts features from such data, saving time and enhancing scalability, versatility, making it suitable for diverse and complex datasets in the future.

\begin{table}[h]
\small
\centering
\caption{Comparative Analysis of Device-type Identification: Hand-crafted Features vs. End-to-end Learning}
\begin{tabular}{c|c|c|c|c|c}
\toprule
\textbf{Features} & \textbf{Dataset} & \textbf{ACC (\%)} & \textbf{REC (\%)} & \textbf{PRE (\%)} & \textbf{F1 (\%)} \\
\midrule
\multirow{3}{*}{\begin{tabular}[c]{@{}c@{}}\textbf{Automatic} \\ \textbf{Features}\end{tabular}} & TMC & 99.99 & 99.99 & 99.99 & 99.99 \\
& IMC & 100 & 100 & 100 & 100 \\
& CIC22 & 99.97 & 99.78 & 99.69 & 99.73 \\
\midrule
\multirow{3}{*}{\begin{tabular}[c]{@{}c@{}} \textbf{Hand-crafted} \\ \textbf{Features}\end{tabular}} & TMC & 99.99 & 99.99 & 99.99 & 99.99 \\
& IMC & 99.99 & 99.99 & 99.99 & 99.99 \\
& CIC22 & 99.99 & 99.99 & 99.99 & 99.99 \\
\bottomrule
\end{tabular}
\label{Table:classification_device}
\end{table}

\subsection{Feature Extraction Time Comparison}
In this section, we investigate the time efficiency of both classical and \gls{dl} approaches, see Table \ref{Table:classification_time}. Our analysis reveals that while hand-crafted features consistently produce accurate results for device-type identification, their extraction from datasets requires a significant time investment, precisely 26 milliseconds (ms) for one packet. This becomes particularly critical in real-world scenarios where factors like time constraints and scalability emerge.

\begin{table}[h]
\small
\centering
\caption{Comparative Analysis of Feature Time Extraction for Device-type Identification Task: Hand-crafted Features vs. End-to-end Learning}
\begin{tabular}{l|cc}
\toprule
\textbf{Methodology} & \textbf{Random Forest} & \textbf{Deep Learning } \\
\midrule
\textbf{Extraction Time (ms)} &  26 &  0.3 \\
\midrule
\textbf{Inference Time (ms)} &   0.001 &  2.6  \\
\midrule
\textbf{Total Time (ms)} & 26.001   &  \textbf{2.9} \\
\bottomrule
\end{tabular}
\label{Table:classification_time}
\end{table}

\section{Limitation and Future Work}

The primary objective of our study was to showcase the effectiveness of an end-to-end learning approach in accurately classifying attacks and device types within IoT environments, as opposed to relying on manually engineered features created by engineers. By achieving this, we have taken a crucial step in encouraging the broader use of \gls{dl} methods in the IoT field, especially given the growing volume of data. However, it is imperative to acknowledge that integrating DL into IoT presents noteworthy limitations. The foremost concern lies in the interpretability of DL models, a challenge that becomes especially critical within IoT scenarios. Furthermore, deploying DL models on resource-constrained IoT devices gives rise to substantial complexities, spanning challenges related to computational resources, energy efficiency, and real-time processing constraints. Additionally, as part of our findings, we highlight the limitations associated with data privacy concerns. Furthermore, investigating strategies for seamlessly integrating and harmonizing datasets originating from a diverse array of IoT devices is imperative.  This approach will allow us to train DL models on more extensive datasets, enhancing their robustness.

\section{Conclusion}
This research highlights the promise of \gls{dl} techniques, specifically end-to-end learning, for vital \gls{iot} tasks like attack classification and device-type identification. By leveraging the power of an end-to-end learning approach, we were able to surpass the limitations of hand-crafted features and achieve comparable or superior results. Our methodology involved training and evaluating a DL model on diverse \gls{iot}-related datasets, providing insights into the effectiveness and adaptability of these models to various datasets. In addition,  It is important to note that although hand-crafted features achieved good results in identifying device-types, they take a long time to process, about 26 milliseconds for just one packet of data against 0.3 milliseconds with \gls{dl}. This time factor is critical in real-world situations where we must consider speed and scalability.

\bibliographystyle{ACM-Reference-Format}
\bibliography{software.bib}

\end{document}